\documentstyle[12pt]{article}
\def\endproof{\hbox to \hsize{\hfil $\Box$}}

\newcommand{\be}{\begin{equation}}
\newcommand{\ee}{\end{equation}}
\newcommand{\bea}{\begin{eqnarray}}
\newcommand{\eea}{\end{eqnarray}}
\def\pd{\partial}
\def\no{\nonumber}

\begin{document}
\hbox to \hsize{\hfil DTP 96/31}
\hbox to \hsize{\hfil July, 1966}
\bigskip

\bigskip
\centerline{{\Large\bf Two-index generalisations of Superconformal Algebras.}}

\vspace{.4in}
\centerline{{\bf D.B. Fairlie}\footnote{e-mail: David.Fairlie@durham.ac.uk}}
\vspace{.2in}
\centerline{Dept. of Mathematical Sciences, University of Durham,}
\vspace{.1in}
\centerline{Durham, DH1 3LE, England}

\vspace{.3in}
\centerline{{\bf J. Nuyts}\footnote{e-mail: Jean.Nuyts@umh.ac.be}}
\vspace{.1in}
\centerline{ Physique Th\'eorique et Math\'ematique, Universit\'e de Mons-Hainaut,}
\vspace{.1in}
\centerline{20 Place du Parc, 7000 Mons, Belgium.}

\begin{abstract}{The superconformal algebras of Ademollo et al are generalised to a multi-index form. The structure obtained is similar to the Moyal Bracket 
analogue of the Neveu-Schwarz Algebra}.
\end{abstract}
\newpage

\section{ Introduction }
For some years now it has been recognised that  the infinite 
classical Lie algebras may all be obtained from some specialisation of the 
Moyal algebra \cite{cos}\cite{nuyt}\cite {fletch}. The supersymmetric
extension in terms of the generalisation to two indices of the Neveu-Schwarz algebra is also known \cite{cos} :
\bea
\label{ns}
{[L_{m n},L_{rs}]}&=&\frac{1}{\sigma}\sin(\sigma (ms-nr))L_{m+r,n+s}\no\\
{[L_{m n},G_{rs}]}&=&\frac{1}{\sigma}\sin(\sigma (ms-nr))G_{m+r,n+s}\\
{\{ G_{m n},G_{rs} \}}&=&\cos(\sigma (ms-nr))L_{m+r,n+s}.\no
\eea
The object of this letter is to discover in what form the superconformal
generalisations of the Virasoro algebra of Ademollo et al \cite{ademollo} can be generalised to a multi-index algebra. First of all we shall remind the reader of the general constraints upon such an algebra, and in particular, the 
equivalence up to renormalisation, which restrict the allowable possibilities.

The construction of a Lie algebra of the form
\be
[L_{mr},L_{ns}]=\lambda(m,r;n,s) L_{m+n,r+s}
\label{(1.1)}
\ee
with, obviously, the antisymmetry
\be
\lambda(m,r;n,s)=-\lambda(n,s;m,r)
\label{(1.2)}
\ee
involves the determination of functions $\lambda(m,r;n,s)$ which fulfill
the Jacobi equations
\be
\sum_{perm}\lambda(m,r;n+p,s+q)\lambda(n,s;p,q)=0,
\label{(1.3)}
\ee
where the sum goes over the three cyclic permutations of  $(m,r)$, $(n,s)$ and
$(p,q)$.

A renormalisation $L^R_{m,r}$ can be performed on $L_{mr}$ with an
arbitrary function $f(m,r)$ \cite{nuyt}
\be
L^R_{mr}=f(m,r)L_{mr}.
\label{(1.4a)}
\ee
The renormalised $\lambda_R$ is
\be
\lambda_R(m,r;n,s)=\lambda(m,r;n,s)R(m,r;n,s),
\label{(1.4b)}
\ee
where the renormalisation factor is
\be
R(m,r;n,s)=\frac{f(m,r)f(n,s)}{ f(m+n,r+s)}.
\label{(1.4c)}
\ee
These renormalisation factors must be taken into consideration to
reduce a given algebra to its simplest form.


Let $\lambda^p(m,r;n,s)$ be the most general polynomial of total
degree $p$ in $m,r,n,s$.
\be
\lambda^p(m,r;n,s)=\sum_{j=0}^p\sum_{k=0}^{p-j}
                   \sum_{l=0}^{p-j-k} 
                    c^p(j,k,l,p-j-k-l)m^j r^k n^l s^{p-j-k-l}.
\label{(1.5)}
\ee
The following results were established using computer algebra.
\subsection{Proposition I}
The most general solution of (\ref{(1.3)}) with a finite polynomial is
\bea
\lambda(m,r;n,s)&=&\no\\
\lambda^2_S(m,r;n,s)&+&\lambda^1_S(m,r;n,s)=l_0(ms-nr)+l_1(m-n)+l_2(r-s).
\label{sol1}
\eea
In the above $l_0,\ l_1,\ l_2$ are arbitrary constants.
\subsection{Proposition (II)}
The function
\be
\lambda_R(m,r;n,s)=l_0 \sin(\sigma(ms-nr))
\label{(1.22)}
\ee
is the unique exact solution of the Jacobi equations,  which extends
polynomially from the quadratic case up to suitable renormalisations.
\vskip 5pt
The important point is that there are no such extensions if linear terms are present in addition to the quadratic ones, contrary to the views implicit in 
\cite{fletch}.
Suppose that we want to find a first order correction in the parameter $\epsilon$ to, say,
$\lambda^1_S(m,r;n,s)$, whose lowest degree $p$ is of the form
$\lambda^p(m,r;n,s)$. More precisely, suppose  
\be
\lambda(m,r;n,s)=\lambda^1_S(m,r;n,s)+\epsilon (\lambda^p(m,r;n,s)+\ldots).
\label{1.12}
\ee
where $\ldots$ stands for terms of degree higher than $p$.
Then defining the generic functions $E(q,p)$ with which the Jacobi identities can be expressed, as
\bea
E(q,p)&=&\sum_{perm}(\lambda^q(m,r;n+k,s+t)\lambda^p(n,s;k,t)\no\\
                      &+&\lambda^p(m,r;n+k,s+t)\lambda^q(n,s;k,t)),
\label{jac}
\eea
it is easy to show that, to first non trivial order in $\epsilon$, $\lambda^p(m,r;n,s)$ has to satisfy 
\be
E(1,p)=0.
\label{jac1}
\ee
Using a renormalisation (see (\ref{(1.4b)})) with $f(m,r)$ of the form
\be
f(m,r)=1-\epsilon g^{p-1}(m,r),
\label{(1.14)}
\ee
with $g^{p-1}(m,r)$ an arbitrary function of $m$ and $s$ of total degree $p-1$,
one sees easily that the correction in $\epsilon$ in (\ref{jac1}) 
is renormalised  to 
\be
\epsilon\lambda^p_R(m,r;n,s)=\epsilon\left(\lambda^p((m,r;n,s)
            -h^{p-1}(m,r;n,s) \lambda^1_S((m,r;n,s)\right),
\label{(1.15)}
\ee
where
 \be
      h^{p-1}(m,r;n,s)=g^{p-1}(m,r)+g^{p-1}(n,s)-g^{p-1}(m+n,r+s).
\label{new}      
\ee

The results can be summarized as follows
\subsection{Proposition (III)}
Except for $p=2$, the most general correction of the form 
(\ref{(1.15)}) to the linear exact solution can be renormalised away to
$\epsilon\lambda^p_R(m,r;)=0$ by
choosing $g^{p-1}(m,r;n,s)$ suitably.
This has been checked explicitly for $3\leq p\leq 7$.
 it follows that the conjecture that ``there does not exist any analytical $\lambda$ whose first
order is linear in its variables'' appears extremely likely.

\section{The generalisation to arbitrary superconformal algebras}
First of all we obtain the most general algebra corresponding to the 
conformal superalgebra found Ademollo et al in the case $ N=2$ under the 
hypothesis that the structure
functions are polynomials of degree at most two in their
variables. This algebra takes the following form, dependent upon one parameter
$p$ 
 found using REDUCE;
\bea\label{one}
{[L_{mr},L_{ns}]}&=&((ms-nr)+p(m-n))L_{m+n,r+s}\no\\
{[L_{mr},T_{ns}]}&=&((ms-nr)-pn) T_{m+n,r+s}\no\\
{[L_{mr},G_{ns}^{\alpha}]}& =& ((ms-nr)+{p(m-2n)\over 2}) G_{m+n,r+s}^{\alpha}\no\\
{[T_{mr},T_{ns}]}&=&0\\
{[T_{mr},G_{ns}^{\alpha}]}&=&\epsilon^{\alpha\beta} G_{m+n,r+s}^{\beta}\no\\
{\{G_{mr}^{\alpha},G_{ns}^{\beta}\}}&=&\delta^{\alpha \beta} L_{m+n,r+s}              -\epsilon^{\alpha\beta} ((ms-nr)+\frac{p(m-n)}{ 2}) T_{m+n,r+s}.\no
 \eea

 Here $\epsilon^{\alpha\beta}$ is the usual two-index antisymmetric symbol.
An equivalent algebra, without the free parameter has appeared recently in a
preprint  by
Buffon et al. \cite{buffon}

The solution corresponding to the Moyal algebra, as found by calculations using REDUCE, is as follows;
\bea\label{two}
{[L_{mr},L_{ns}]}&=&\sin(ms-nr)L_{m+n,r+s}\no\\
{[L_{mr},T_{ns}]}&=& \sin(ms-nr) T_{m+n,r+s}\no\\
{[L_{mr},G_{ns}^{\alpha}]}&
      =& \sin(ms-nr) G_{m+n,r+s}^{\alpha}\no\\
{[T_{mr},T_{ns}]}&=&\sin(ms-nr)L_{m+n,r+s}\\
{[T_{mr},G_{ns}^{\alpha}]}&
       =&\epsilon^{\alpha\beta} \cos(ms-nr)G_{m+n,r+s}^{\beta}\no\\
{\{G_{mr}^{\alpha},G_{ns}^{\beta}\}}&
         =&\delta^{\alpha \beta}\cos(ms-nr) L_{m+n,r+s}              
    -\epsilon^{\alpha\beta} \sin(ms-nr) T_{m+n,r+s}.\no  
\eea
An  inessential parameter $\sigma$ (as in (\ref{ns})) has been set to unity.

\section{Extension for $N=4$}
 The $N=4$ case (the Greek indices below run over the values from $1$ to $4$) may be treated in a similar fashion. With $L,\ U,\ G^\mu,\ Q^\mu$ and $A^{\mu\nu}$ behaving as a scalar, pseudoscalar, vector, axial vector and tensor respectively under the action of the orthogonal group, the following solution of the 86 identities which
are a consequence of the 35 Jacobi identities of the problem were found, using REDUCE;
introducing the abbreviations
\bea
S&=&\sin{(\overrightarrow m\wedge \overrightarrow n)} \nonumber\\
C&=&\cos{(\overrightarrow m\wedge \overrightarrow n)},   
\label{8.4}
\eea
where $\overrightarrow m,\ \overrightarrow n$ are
real $2k$-dimensional (or $2k+1$-dimensional) vectors and $\overrightarrow m\wedge \overrightarrow n$
signifies
\be
\overrightarrow m \wedge \overrightarrow n= \sigma_{1,2}(m_1n_2-m_2n_1)+\dots+\sigma_{2k-1,2k}(m_{2k-1}n_{2k}-m_{2k}n_{2k-1}),
\label{def}
\ee
with the coefficients $\sigma_{i,i+1}$ being treated as constant parameters,
one finds the following algebra;
\bea
{[L_{\overrightarrow m},L_{\overrightarrow n}]}
        & =&S L_{\overrightarrow{m}+\overrightarrow{n}}\no\\
{[L_{\overrightarrow m},G^{\mu}_{\overrightarrow n}]}
          &=&S G^{\mu}_{\overrightarrow{m}+\overrightarrow{n}}\no\\
{[L_{\overrightarrow m},A^{\mu\nu}_{\overrightarrow n}]}
          &=&S A^{\mu\nu}_{\overrightarrow{m}+\overrightarrow{n}}  \no \\ 
{[L_{\overrightarrow m},Q^{\mu}_{\overrightarrow n}]}
           &=&S Q^{\mu}_{\overrightarrow{m}+\overrightarrow{n}} \no  \\ 
{[L_{\overrightarrow m},U_{\overrightarrow n}]}
           &=&S U_{\overrightarrow{m}+\overrightarrow{n}}  \no \\ 
{\{G^{\mu}_{\overrightarrow m},G^{\nu}_{\overrightarrow n}\}}
        &=&C \delta^{\mu\nu}L_{\overrightarrow{m}+\overrightarrow{n}} 
         +S A^{\mu\nu}_{\overrightarrow{m}+\overrightarrow{n}}  \no \\ 
{[G^{\mu}_{\overrightarrow m},A^{\nu\rho}_{\overrightarrow n}]}
        &=&C (\delta^{\mu\nu}
               G^{\rho}_{\overrightarrow{m}+\overrightarrow{n}}
           -\delta^{\mu\rho}
               G^{\nu}_{\overrightarrow{m}+\overrightarrow{n}}) 
         +S \epsilon^{\mu\nu\rho\sigma}
               Q^{\sigma}_{\overrightarrow{m}+\overrightarrow{n}}  \no \\ 
{\{G^{\mu}_{\overrightarrow m},Q^{\nu}_{\overrightarrow n}\}}
        &=&-S \delta^{\mu\nu}U_{\overrightarrow{m}+\overrightarrow{n}} 
     +{1\over 2} C \epsilon^{\mu\nu\rho\sigma}
               A^{\rho\sigma}_{\overrightarrow{m}+\overrightarrow{n}}  \no \\ 
{[G^{\mu}_{\overrightarrow m},U_{\overrightarrow n}]}
        &=&- C Q^{\mu}_{\overrightarrow{m}+\overrightarrow{n}}      \\
{[A^{\mu\nu}_{\overrightarrow m},A^{\rho\sigma}_{\overrightarrow n}]}
        &=&- S (\delta^{\mu\sigma}\delta^{\nu\rho}
                  -\delta^{\mu\rho}\delta^{\nu\sigma})
                              L_{\overrightarrow{m}+\overrightarrow{n}}\no\\
         &\ & + C (\delta^{\mu\sigma}
               A^{\nu\rho}_{\overrightarrow{m}+\overrightarrow{n}}
            +\delta^{\nu\rho}
               A^{\mu\sigma}_{\overrightarrow{m}+\overrightarrow{n}}\no\\
               &-&\delta^{\mu\rho}
               A^{\nu\sigma}_{\overrightarrow{m}+\overrightarrow{n}}
               - \delta^{\nu\sigma}
               A^{\mu\rho}_{\overrightarrow{m}+\overrightarrow{n}})\no\\
         &\ & + S \epsilon^{\mu\nu\rho\sigma}
               U_{\overrightarrow{m}+\overrightarrow{n}} 
 \no  \\ 
{[A^{\mu\nu}_{\overrightarrow m},Q^{\rho}_{\overrightarrow n}]}
        &=&- S \epsilon^{\mu\nu\rho\sigma}
               G^{\sigma}_{\overrightarrow{m}+\overrightarrow{n}}
         + C (\delta^{\mu\rho}
               Q^{\nu}_{\overrightarrow{m}+\overrightarrow{n}}
                  -\delta^{\nu\rho}
               Q^{\mu}_{\overrightarrow{m}+\overrightarrow{n}})  \no \\ 
{[A^{\mu\nu}_{\overrightarrow m},U_{\overrightarrow n}]}
     &=&{1\over 2} S \epsilon^{\mu\nu\rho\sigma}
               A^{\rho\sigma}_{\overrightarrow{m}+\overrightarrow{n}}
                         \no  \\
{\{Q^{\mu}_{\overrightarrow m},Q^{\nu}_{\overrightarrow n}\}}
        &=& C \delta^{\mu\nu}L_{\overrightarrow{m}+\overrightarrow{n}} 
         + S A^{\mu\nu}_{\overrightarrow{m}+\overrightarrow{n}} \no  \\ 
{[Q^{\mu}_{\overrightarrow m},U_{\overrightarrow n}]}
     &=& C G^{\mu}_{\overrightarrow{m}+\overrightarrow{n}}
                          \no \\
{[U{\overrightarrow m},U_{\overrightarrow n}]}
     &=&  S L_{\overrightarrow{m}+\overrightarrow{n}}.\no             
\label{three}
\eea
In the anticommutator $\{G,Q\}$ and the commutators $[G,U]$ and $[Q,U]$,
one might have expected $C$ where $S$ appears and vice versa, especially as the
standard $N=4$ Ademollo algebra is not recovered by a limiting procedure.  

However, there is a representation of this algebra (for two-vector indices) in terms of functions $F(x,y)_{mn}$ which act as basis functions for the Star product,  tensored with a Clifford algebra. The motivation which lies behind  this representation comes from the precise behviour of the generators under the orthogonal group. This procedure obviously generalises to even $N$.
 The Star product of two functions
$F(x,y)$ and $G(x,y)$ is the associative product
\be
F(x,y)\star G(x,y) = \lim_{\stackrel{x'\rightarrow x}{y'\rightarrow y}}
\exp\left(\frac{\pd}{\pd x}\frac{\pd}{\pd y'}-\frac{\pd}{\pd x'}\frac{\pd}{\pd y}\right)F(x,y)G(x',y').
\label{star}
\ee
Choose $F(x,y)_{mn}=\exp(mx+iny)$ and represent the operators as follows, in terms of the a direct product of the familiar 4 dimensional Dirac matrices $\gamma^\mu,\ \gamma^5=\gamma^0\gamma^1\gamma^2\gamma^3$ and the functions $F_{mn}$; 
\bea\label{cliff}
L_{mn}&=&1\!\!1\otimes F_{mn}\no\\
         G_{mn}^\mu&=&\gamma^\mu \otimes F_{mn}\no\\
A_{mn}^{\mu,\nu}&=&{1\over2}(\gamma^\mu\gamma^\nu-\gamma^\nu\gamma^\mu) \otimes F_{mn}\\
          Q_{mn}^\mu&=&\gamma^5\gamma^\mu \otimes F_{mn}\no\\
U_{mn}&=&\gamma^5\otimes F_{mn}.\no
\eea
These assignments provide a representation of the algebra (\ref{three}) where the
product is used in the sense of the Star product. In the case where $\lambda={2\pi\over{\rm integer}}$, $F_{m,n}$ can be represented by matrices in the Weyl representation of a unitary algebra, and the product is then ordinary multiplication. 
\section{Conclusion}
It is found that the generalisation of infinite superconformal algebras 
to multi-index algebras is of surprisingly restrictive type, being patterned after the Neveu-Schwarz example as
a direct product of a Clifford algebra with a Star product.

\section{Acknowledgements}
Both the authors are grateful  to the E.C. fund  ERB-CHR-XCT 920069
for the opportunity to visit each other at our home institutions, where this work was initiated.
One of them, (J.N.) would like to thank professors 
S. Randjbar-Daemi and F. Hussain for hospitality at the I.C.T.P. 
(Trieste) where this work was completed.

We are both indebted to Cosmas Zachos for several illuminating discussions.

\end{document}